\documentclass[twocolumn,showpacs,preprintnumbers,amsmath,amssymb,prd]{revtex4}
\usepackage{graphicx}
\usepackage{bm}

\begin{document}
\preprint{PNU-NTG-09/2004}
\preprint{PNU-NURI-08/2004}
\preprint{BNL-NT-04/20}
\preprint{TPJU-7/2004}

\title{Magnetic moments of the exotic pentaquark baryons within the
chiral  quark-soliton model}\thanks{A talk presented at the
YITP workshop held in Kyoto, 17 - 19, Feb., 2004.
}
\author{Hyun-Chul Kim}
\email{hchkim@pusan.ac.kr}
\affiliation{Department of Physics, and \\
Nuclear Physics and Radiation Technology Institute (NuRI),
Pusan National University, 609-735 Busan, \\ Republic of Korea}
\author{Micha{\l} Prasza{\l}owicz}
\email{michal@quark.phy.bnl.gov}
\affiliation{Nuclear Theory Group,
Brookhaven National Laboratory,
Upton, NY 19973-5000,
and \\
 M. Smoluchowski
Institute of Physics, Jagellonian University, Krak{\'o}w, Poland}
\date{May 2004}

\begin{abstract}
We present in this talk recent results of the magnetic moments of the
baryon antidecuplet within the framework of the chiral quark-soliton
model in the chiral limit.  The dynamic parameters of the model are
fixed by using the experimental data for those of the baryon octet.
Sum rules for the magnetic moments are derived.  We found that the
magnetic moments of the baryon antidecuplet have opposite signs to
their charges.  The magnetic moments of the neutral baryon
antidecuplet turn out to be compatible with zero.
\end{abstract}
\maketitle

\section{Introduction}
The LEPS collaboration~\cite{Nakano:bh} announced the finding of
the $\Theta^+$ consisting of four quarks and one anti-quark
($\mathrm{uudd}\bar{\mathrm{s}}$), motivated by a theoretical
prediction of the chiral soliton model~\cite{Diakonov:1997mm}.
Since then, a great amount of experimental and theoretical
works~\cite{Praszalowicz:2003ik,Stepanyan:2003qr,Barmin:2003vv,
Barth:2003es,Aktas:2004qf,Chekanov:2004kn,neutrino,NA49,quark,chiral,
chiral2,lattice,bound,Stancu:2003if,Sugiyama:2003zk,Liu:2003ab,
Zhu:2003ba,summag,groupth,Hong:2004xn,Praszalowicz:2004dn} has
been published.

In order to describe the photo-production of pentaquark
baryons~\cite{Liu:2003rh,Nam:2003uf,Oh:2003kw}, we
need information on their magnetic moments.  Since there is no
experimental data available, one needs a theoretical guideline to
estimate them.  Recently, the present authors investigated the
magnetic moments of the pentaquark baryons within the chiral
quark-soliton model in a ``{\em
  model-independent}'' way~\cite{Adkins:cf,Kim:1997ip,Kim:1998gt}.
In this approach the
dynamical model parameters are fixed by using the experimental data
of the octet magnetic moments~\cite{Kim:2003ay}.  However,
not all parameters can be constrained that way. Hence
Ref.~\cite{Kim:2003ay} used some additional information
based on the dynamical model calculations. Therefore
the analysis of Ref.~\cite{Kim:2003ay}  was not
self-consistent.  In this talk, we will present the recent results for
the magnetic moments of the baryon antidecuplet in the chiral limit
with the complete set of parameters fixed from the experimental
data.

\section{Formalism}
The magnetic moments of the baryon antidecuplet can be
defined as the following one-current baryon matrix element:
\begin{equation}
\label{mat}\langle B_{\overline{10}} | \bar{\psi}(z)\gamma_{\mu}\hat{Q}
\psi(z) | B_{\overline{10}} \rangle,
\end{equation}
where $\hat{Q}$ denotes the charge operator of quarks in SU(3) flavor space,
defined by
\begin{equation}
\label{Eq:charge}\hat{Q}\;=\; \left(
\begin{array}{ccc}
\frac23 & 0 & 0 \\
0 & -\frac13 & 0 \\
0 & 0 & -\frac13
\end{array}
\right)\;=\; \frac12 \left(\lambda^3 + \frac{1}{\sqrt{3}}\lambda^8\right).
\end{equation}
In the nonrelativistic limit, the Sachs form factors
$G_E$ and $G_M$ can be related to the time and space components
of the $\mbox{U}_V(3)$ vector currents, respectively:
\begin{equation}
\langle B_{\overline{10}}(p') | \bar{\psi}(z)\gamma_{0}\hat{Q}
\psi(z) | B_{\overline{10}}(p)\rangle
= G_E^{B_{\overline{10}}}
(Q^2),
\end{equation}
\begin{eqnarray}
&&\langle B_{\overline{10}}(p') | \bar{\psi}(z)\gamma_{i}\hat{Q}
\psi(z)|B_{\overline{10}}(p)\rangle \cr
&=& \frac{1}{2M_N}
G_M^{B_{\overline{10}}} (Q^2) i\epsilon_{ijk} q^j \langle s' |
\sigma_k |s\rangle,
\label{Eq:form}
\end{eqnarray}
where $\sigma_k$ denotes Pauli spin matrices while $|s\rangle$ is
the corresponding spin state of the baryon. The magnetic moments
$\mu_{B_{\overline{10}}}$ corresponding to the vector currents are
identified with $G^{B_{\overline{10}}}_{M} (0)$.

The collective magnetic moment operator can be obtained
schematically by differentiating the effective chiral action with
the external source corresponding to the magnetic moments as
follows:
\begin{eqnarray}
\hat{\mu}_k &=& \frac{\delta}{\delta s_k} S_{\rm eff} \cr
&=&-N_c \frac{\delta }{\delta s_k} {\rm Tr}\log\left[
i\partial _{4} + iH(U_c^{\gamma_5})-\Omega + i\gamma _{4}
R^\dagger \hat{m} R \right.\cr
&&\left.-i s_m \epsilon_{ilm}\gamma_4 \gamma_ix_l R^\dagger\hat{Q}R
\right],
\label{eq:mhat}
\end{eqnarray}
where $H(U^{\gamma_5})$ is the one-body Dirac Hamiltonian defined by
\begin{equation}
 H(U^{\gamma_5}) = \frac{\bm{\alpha}\cdot\bm{\nabla}}{i} + \beta M
 U^{\gamma_5} \label{eq:ham}.
\end{equation}
$U^{\gamma_5}$ stands for the chiral soliton field:
\begin{equation}
U^{\gamma_5} = \frac{1+\gamma_5}{2} U + \frac{1-\gamma_5}{2} U^\dagger
\end{equation}
with the trivial embedding
\begin{equation}
 U_c = \left(\begin{array}{cc} U_{\rm SU(2)}& 0 \\ 0& 1
 \end{array}\right).
\end{equation}
Here, $U(SU(2))$ denotes the SU(2) soliton field.  $\Omega=\frac12
\lambda_a\Omega_a$ in
Eq.(\ref{eq:mhat}) designates the angular velocity of the soliton,
which is related to the right angular momentum operator after the
zero-mode quantization:
\begin{equation}
{\cal R}_a \;=\; -\Omega_b I_{ba} + \frac{N_c}{2\sqrt{3}}\delta_{8a}
- 2K_{ab} m_8 D^{(8)}_{8b}(R)
\end{equation}
with the current quark masses
\begin{equation}
 \hat{m} = \left(\begin{array}{ccc} m_{\rm u} & 0& 0 \\ 0& m_{\rm d}
    & 0 \\ 0&0 & m_{\rm s} \end{array}\right) = m_0{\bm 1} + m_8
    \lambda_8.
\end{equation}
$I_{ab}$ and $K_{ab}$ stand for the moments of inertia.  Taking into
account the rotational $1/N_c$ corrections as well as the linear
$m_{\rm s}$ corrections, we arrive at the following collective
operator of the magnetic moments:
\begin{eqnarray}
\hat \mu_3  &=& w_1
\;D_{Q3}^{(8)}\;+\;w_2d_{pq3}D_{Qp}^{(8)} \hat S_q\;+
\;\frac{w_3}{\sqrt{3}}D_{Q8}^{(8)} \hat S_3
 \nonumber \\
&+& m_{\rm s}\left[\frac{w_4}{\sqrt{3}}d_{pq3}D_{Qp}^{(8)}D_{8q}^{(8)}
\right. \cr
&+& w_5\left(D_{Q3}^{(8)}D_{88}^{(8)}+D_{Q8}^{(8)}D_{83}^{(8)}\right) \cr
&+&\left. w_6\left(
    D_{Q3}^{(8)}D_{88}^{(8)}-D_{Q8}^{(8)}D_{83}^{(8)}\right)
\right] .
\end{eqnarray}
where the dynamical variables $w_{i}$ contain information of the
dynamics of the chiral soliton, which are independent of baryons
considered.  They can be generically written
in terms of the inertia parameters of the soliton in the $\chi$QSM:
\begin{equation}
\sum_{m,n}\langle n|\Gamma_{1}|m\rangle\langle m|\Gamma_{2}|n\rangle
\mathcal{R}(E_{n},E_{m},\Lambda), \label{spec}%
\end{equation}
where $\Gamma_{i}$ denote spin-isospin operators acting on the quark eigenstates
$|n\rangle$ of the one-body Dirac Hamiltonian~(\ref{eq:ham}) in the
soliton-background field.  The double sum over all the eigenstates can
be evaluated
numerically~\cite{Goeke:fk,Wakamatsu:1990ud,Blotz:1992pw}.  Since its
sea part diverges, we need the regularization expressed by
$\mathcal{R}$ with the cut-off parameter $\Lambda$ fixed to the pion
decay constant.  In this work, we will not calculate the dynamical
variables $w_{i}$ numerically but we
will constrain them using the experimental data of
the octet magnetic moments~\cite{Kim:1997ip,Kim:1998gt}.
$D_{ab}^{(\mathcal{R})}(R)$ stands for the SU(3) Wigner function,
$R(t)$ is the time-dependent SU(3) matrix responsible for the rotation
of the soliton in the collective
coordinate space \cite{Blotz:1992pw,Christov:1995vm}.
$\hat{J}_{a}$ denotes an operator of the generalized spin acting on
the baryonic wave functions $\psi_{B_{\mathcal{R}}}(R)$.
Since the SU(3) symmetry breaking introduce the mixing of the pure
antidecuplet states with higher representations, we have to calculate
the wave-function corrections.  However, we will present here only the
results of the magnetic moments in the chiral limit, namely, we will
put $m_{\rm s}=0$ in Eq.(\ref{eq:mhat}).

In order to evaluate the magnetic moments of the baryon antidecuplet,
we have to sandwich the collective opreator in Eq.(\ref{eq:mhat})
between the collective baryon $\overline{\bm 10}$ states:
\begin{equation}
\mu_{B_{\overline{10}}}=\int dR\psi_{B_{\overline{10}}}^{\ast}(R)\hat{\mu
}(R)\psi_{B_{\overline{10}}}(R),\label{Eq:matrix_e}%
\end{equation}
where the collective wave functions $\psi_{B_{\mathcal{R}}}(R)$ are
defined as follows:
\begin{equation}
\psi_{B_{\mathcal{R}}}(R)
=\sqrt{\mathrm{dim}(\mathcal{R})}(-1)^{J_{3}-Y^{\prime
}/2}D_{Y,T,T_{3};Y^{\prime},J,-J_{3}}^{(\mathcal{R})\ast}(R).\label{Eq:wave_f}%
\end{equation}
Here $\mathcal{R}$ stands for the allowed irreducible representations
of the SU(3) flavor group, \emph{i.e.}
$\mathcal{R}=8,10,\overline{10},\cdots$ and $Y,T,T_{3}$ are the
corresponding hypercharge, isospin, and its third component,
respectively. Right hypercharge $Y^{\prime}$ is constrained to be
unity for the physical spin states for which $J$ and $J_{3}$ are spin
and its third component. Note that under the action of left (flavor)
generators
$\hat{T}_{\alpha}=-D_{\alpha\beta}^{(8)}\hat{J}_{\beta}$
$\psi_{B_{\mathcal{R}}}$ transforms
like a tensor in representation $\mathcal{R}$, while under the right
generators $\hat{J}_{\alpha}$ like a tensor in $\mathcal{R}^{\ast}$ rather
than $\mathcal{R}$. This is the reason why operators like the one multiplied
by $w_{2}$ in Eq.(\ref{eq:mhat}) have different matrix elements for the decuplet
(which is spin $3/2$) and antidecuplet (which is spin $1/2$). The other two
operators multiplied by $w_{1,3}$ have the same matrix elements between
decuplet and antidecuplet states. 

The matrix elements of Eq.(\ref{Eq:matrix_e}) are expressed  in
terms of SU(3) Clebsch-Gordan coefficients \cite{KW}.  Having
scrutinized the results, we find the following simple expression:
\begin{eqnarray}
\mu_{B_{\overline{10}}}&=&-\frac{1}{12}\,\left(w_{1}+\frac{5}{2}w_{2}
-\frac{1}{2}w_{3}\right)
Q_{B_{\overline{10}}}\,J_{3},\label{m0B10},\\
\mu_{B_{10}}&=&-\frac{1}{12}\,\left(w_{1}-\frac{1}{2}w_{2}
-\frac{1}{2}w_{3}\right)
Q_{B_{10}}\,J_{3},\label{mB10}
\end{eqnarray}
where $Q_{B_{\overline{10}}}$ is the charge of the antidecuplet expressed by the
Gell-Mann--Nishijima relation:
\begin{equation}
Q_{B_{\overline{10}}}=T_{3}+\frac{Y}{2}.
\end{equation}
$J_{3}$ is the corresponding third component of the spin.

\section{Numerical fits}
In order to fit the parameters $w_{i}$, it is convenient to introduce two
parameters consisting of $w_{1}$, $w_{2}$ and $w_{3}$:
\begin{equation}
v=\frac{1}{60}\left(  w_{1}-\frac{1}{2}w_{2}\right)  ,~~w=\frac{1}{120}%
\;w_{3}. \label{vw}%
\end{equation}
In Ref.~\cite{Kim:1997ip} the octet and decuplet magnetic moments were
expressed as follows:
\begin{align}
\mu_{p}  &  =\mu_{\Sigma^{+}}=-8v+4w,\nonumber\\
\mu_{n}  &  =\mu_{\Xi^{0}}=6v+2w,\nonumber\\
\mu_{\Lambda}  &  =-\mu_{\Sigma^{0}}=3v+w,\nonumber\\
\mu_{\Sigma^{-}}  &  =\mu_{\Xi^{-}}=2v-6w,\nonumber\\
\mu_{B_{10}}  &  =\frac{15}{2}\left(  -v+w\right)
\,Q_{B_{10}}\label{Eq:all_mom}.
\end{align}
which are in fact the well-known SU(3) formulae for the magnetic moments.

On the other hand, magnetic moments of the baryon antidecuplet
(\ref{m0B10}) can be rewritten as:
\begin{equation}
\mu_{B_{\overline{10}}}=\left[  \frac{5}{2}\left(  -v+w\right)  -\frac{1}%
{8}w_{2}\,\right]  Q_{B_{\overline{10}}}.\label{Eq:final1}%
\end{equation}
Equation (\ref{Eq:final1}) is different from the decuplet in
Eq.(\ref{mB10}) by the second term proportional to $w_2$.  The factor
three difference in the first term between Eq.(\ref{Eq:all_mom}) and
Eq.(\ref{Eq:final1}) is due to the fact that the baryon antidecuplet
has spin $1/2$, while the decuplet has $3/2$.

Using Eq.(\ref{Eq:final1}), we are able to derive the sum rules which
are similar to the generalized Coleman and Glashow sum
rules~\cite{Coleman:1961jn} in the chiral limit:
\begin{align}
\mu_{\Sigma_{\overline{10}}^{0}}  &  =\frac{1}{2}\left(  \mu_{\Sigma
_{\overline{10}}^{+}}+\mu_{\Sigma_{\overline{10}}^{-}}\right)  ,\nonumber\\
\mu_{\Xi_{3/2}^{+}}+\mu_{\Xi_{3/2}^{--}}  &  =\mu_{\Xi_{3/2}^{0}}+\mu
_{\Xi_{3/2}^{-}},\nonumber\\
\sum\mu_{B_{\overline{10}}}  &  =0. \label{Eq:ColGl}%
\end{align}

As discussed in Ref.~\cite{Kim:1997ip}, there are different ways to fix the
parameters $v$ and $w$ by using the experimental data of the octet magnetic
moments. Here, we simply fit the proton and neutron magnetic moments (fit I):
\begin{equation}%
\begin{array}
[c]{llllr}%
v & = & (2\mu_{\mathrm{n}}-\mu_{\mathrm{p}})/20 & = & -0.331,\\
w & = & (4\mu_{\mathrm{n}}+3\mu_{\mathrm{p}})/20 & = & 0.037,
\end{array}
\label{Eq:fitI}%
\end{equation}
and use the following ''average'' values (fit II):
\begin{eqnarray}%
v & = & \left(  2\mu_{\mathrm{n}}-\mu_{\mathrm{p}}+3\mu_{\Xi^{0}}+\mu_{\Xi
^{-}}-2\mu_{\Sigma^{-}}-3\mu_{\Sigma^{+}}\right)  /60 \cr
& = & -0.268,\\
w & = & \left(  3\mu_{\mathrm{p}}+4\mu_{\mathrm{n}}+\mu_{\Xi^{0}}-3\mu
_{\Xi^{-}}-4\mu_{\Sigma^{-}}-\mu_{\Sigma^{+}}\right)  /60 \cr
& = & 0.063.
\label{Eq:mean}%
\end{eqnarray}
to fix parameters $v$ and $w$. It was shown in Ref.~\cite{Kim:1997ip} that
combinations of Eq.(\ref{Eq:mean}) are independent of the linear corrections
due to the nonzero strange quark mass $m_{\mathrm{s}}$.  Thus, fit II
is also valid when the SU(3)-symmetry breaking is taken
into account, while fit I will be changed
by the corrections of order $\mathcal{O}(m_{\mathrm{s}})$.

While $v$ and $w$ can be fixed by Eq.(\ref{Eq:mean}), we are not
able to get $w_2$ from the analysis in the chiral limit.  In order
to fix it, we have to carry out the full analysis with the $m_{\rm
s}$ corrections considered in Ref.~\cite{Yangetal}. Otherwise, we
have to take it from the model calculation~\cite{Kim:2003ay}.  The
value of $w_2$ obtained with the SU(3) symmetry breaking depends on
the pion-nucleon $\Sigma$ term \cite{Praszalowicz:2004dn}, and for the
values of $\Sigma\sim 70$~Mev we get: 
\begin{equation}
m_s \; w_2 = 9.81.
\end{equation}
Compared to the value from the model $m_s \;w_2^{\chi {\rm QSM}} \sim 5$
used in Ref.~\cite{Yangetal}, it is almost two times larger.

\section{Results and Discussion}
The results of these fits are listed in Table I.
\begin{table}[h]
  \centering
  \begin{tabular}{lrrrr}
& exp. & fit I & fit II & $\chi$QSM \\\hline
$p$ & $2.79$ & input & $2.39$ & $2.27$\\
$n$ & $-1.91$ & input & $-1.49$ & $-1.55$\\
$\Lambda$ & $-0.61$ & $-0.96$ & $-0.74$ & $-0.78$\\
$\Sigma^{+}$ & $2.46$ & $2.79$ & $2.38$ & $2.27$\\
$\Sigma^{0}$ & $(0.65)$ & $0.96$ & $0.74$ & $0.78$\\
$\Sigma^{-}$ & $-1.16$ & $-0.89$ & $-0.90$ & $-0.71$\\
$\Xi^{0}$ & $-1.25$ & $-1.91$ & $-1.49$ & $-1.55$\\
$\Xi^{-}$ & $-0.65$ & $-0.89$ & $-0.90$ & $-0.71$\\\hline
$\Delta^{++}$ & $4.52$ & $5.52$ & $4.92$ & $4.47$\\
$\Omega^{-}$ & $-2.02$ & $-2.76$ & $-2.46$ & $-2.23$\\\hline
$\Theta^{+}$ & $?$ & $-0.31$ & $-0.40$ & $0.12$ \\
$\Xi^{--}$ & $?$ & $0.62$ & $0.8$ & $-0.24$ \\
  \end{tabular}
  \caption{The magnetic moments of the baryon octet, decuplet, and
    antidecuplet in the chiral limit.  The experimental value for the
    $\Delta^{++}$ magnetic moments is taken from Ref.\cite{Deltapp2}.}
  \label{tab:1}
\end{table}
We see that the quality of these fits is rather poor reaching in its
worst case about $25\%$ accuracy, which indicates the importance of the
SU(3)-symmetry breaking corrections.  The results for the baryon
antidecuplet are rather remarkable, because their magnetic moments have
opposite signs to their charges.  Since in the chiral limit
the decuplet and antidecuplet
magnetic moments are proportional to their charges, those for the
neutral baryons turn out to be zero.

If in the $\chi$QSM one artificially sets the soliton size
$r_{0}\rightarrow0$, then the model reduces to the free valence quarks which,
however, ''remember'' the soliton structure. In this limit, many
quantities, for example the axial-vector couplings, are given as
ratios of the group-theoretical factors \cite{limit}. In the case of
magnetic moments the pertinent expressions are given as a product of
the group-theoretical factor and the model-dependent integral which
we shall in what follows denote by $K$~\cite{paradox}.

Constants $w_{1,2,3}$ entering Eq.(\ref{eq:mhat}) are expressed in
terms of the inertia parameters in the following way:
\begin{equation}
w_{1}=M_{0}-\frac{M_{1}^{(-)}}{I_{1}^{(+)}},\quad w_{2}=-2\frac{M_{2}^{(+)}%
}{I_{2}^{(+)}},\quad w_{3}=-2\frac{M_{1}^{(+)}}{I_{1}^{(+)}}.
\end{equation}
For the soliton size $r_{0}\rightarrow0$ we have \cite{paradox}:%
\begin{eqnarray}
M_{0}&\rightarrow&-2K\ ,\quad\frac{M_{1}^{(-)}}{I_{1}^{(+)}}\rightarrow
\frac{4}{3}K, \cr
\frac{M_{2}^{(+)}}{I_{2}^{(+)}}&\rightarrow&-\frac{4}{3}K
~~~~~~\frac{M_{1}^{(+)}}{I_{1}^{(+)}}\rightarrow-\frac{2}{3}K\ ,
\end{eqnarray}
which give%
\begin{equation}
v=-\frac{7}{90}K,~~w=\frac{1}{90}K,\quad w_{3}=\frac{4}{3}K,
\end{equation}
yielding the magnetic moments of the proton and neutron as follows:
\begin{equation}
\mu_{p}=\frac{2}{3}K,\quad\mu_{n}=-\frac{4}{9}K. \label{Eq:Knp}%
\end{equation}
Hence, the ratio of the proton magnetic moment to the neutron one
takes the value from the nonrelativistic quark model:
\begin{equation}
\frac{\mu_{p}}{\mu_{n}}=-\frac{2}{3}.
\end{equation}
We get for the antidecuplet magnetic moments:
\begin{equation}
\mu_{B_{\overline{10}}}=-\frac{1}{3}KQ_{B_{\overline{10}}}%
\end{equation}
which agrees in sign with the phenomenological value of Table I (note that
$K$ is positive in view of Eq.(\ref{Eq:Knp})). Extracting $K$ from
proton or neutron magnetic moments we get $K=3.4$ and $4.3$ respectively.
These values lead to rather large, bur negative, value of $\mu_{\Theta^+}
=-1.15 \sim -1.4 $ respectively.

\section{Summary}
In the present talk, we determined the magnetic moments of the
positive parity
baryon antidecuplet in a ``{\em model independent}''
analysis, based on the chiral
quark-soliton model in the chiral limit.  Starting from the collective
operators with dynamical parameters fixed by experimental data, we
were able to obtain the magnetic moments of the baryon antidecuplet.
The expression for the magnetic moments of the antidecuplet is
different from those of the baryon decuplet.  We found that the
magnetic moment of $\mu_{\Theta^{+}}$ is about $-0.3 \sim
-0.4\,\mu_{N}$ which differs from the recent results of
Refs.~\cite{summag,Hong:2004xn,Liu:2003ab} and our previous estimate
\cite{Kim:2003ay} where generally $\mu_{\Theta^{+}}$ is small
and positive.

In the present talk, we have presented results
in the chiral limit.  The SU(3)-symmetry
breaking effects will definitely make the magnetic moments of the baryon
antidecuplet deviate from those of the present paper.  There are two different
sources of the SU(3)-symmetry breaking effects: one comes from the collective
operator, the other arises from the fact that the collective wave
functions of the baryon antidecuplet are mixed with the octet,
eikosiheptaplet ({\bf 27}), and $\overline{\bm{ 35}}$ representations.
Moreover, nonanalytical
symmetry breaking effects are of  importance \cite{chiloop}.  The
effect of the SU(3)-symmetry breaking on the magnetic moments of the
antidecuplet baryons has been studied and the results will soon
appear~\cite{Yangetal}.

\section*{Acknowledgments}

H.-Ch.K is grateful to J.K. Ahn (LEPS collaboration), K. Goeke,
A. Hosaka, P.V. Pobylitsa, M.V. Polyakov, G.S. Yang,  and I.K. Yoo
(NA49 collaboration) for valuable discussions.  The present work is
supported by Korea Research Foundation Grant: KRF-2003-041-C20067
(H.-Ch.K.) and by the Polish State Committee
for Scientific Research under grant 2 P03B 043 24 (M.P.). This
manuscript has been authored  under Contract No. DE-AC02-98CH10886
with the U.S. Department of Energy.

\end{document}